\documentclass[manuscript]{aastex}

\def\Msun{\hbox{M$_{\odot}$}}

\shorttitle{FAST X-RAY/IR CROSS-CORRELATIONS AND RELATIVISTIC JET FORMATION IN GRS 1915+105}
\shortauthors{Lasso-Cabrera \& Eikenberry}

\begin{document}


\title{FAST X-RAY/IR CROSS-CORRELATIONS AND RELATIVISTIC JET FORMATION IN GRS 1915+105}


\author{Lasso-Cabrera, N. M.}
\affil{Departamento de Astronom\'ia, Universidad de Concepci\'on, Chile}
\affil{Department of Astronomy, University of Florida, Gainesville, FL 32611}

\and

\author{Eikenberry, S. S.}
\affil{Department of Astronomy, University of Florida, Gainesville, FL 32611}
\email{nestor@astro-udec.cl}



\begin{abstract}
We present cross-correlation analyses of simultaneous X-ray and near-infrared (near-IR) observations of the microquasar GRS 1915+105 during relativistic jet-producing epochs (X-ray class $\alpha$ and $\beta$). While previous studies have linked the large-amplitude IR flares and X-ray behaviors to jet formation in these states, our new analyses are sensitive to much lower-amplitude IR variability, providing more sensitive probes of the jet formation process. The X-ray to IR cross-correlation function (CCF) shows significant correlations which vary in form between the different X-ray states. During low/hard dips in both classes, we find no significant X-ray/IR correlation. During high-variability epochs, we find consistently significant correlations in both $\alpha$ and $\beta$ classes, but with strong differences in the CCF structure. The high-variability $\alpha$ CCF shows strong anti-correlation between X-ray/IR, with the X-ray preceding the IR by $\sim$ 13 $\pm$ 2s. The high-variability $\beta$ state shows time-variable CCF structure, which is statistically significant but without a clearly consistent lag. Our simulated IR light curves, designed to match the observed CCFs, show variably-flickering IR emission during the class $\beta$ high-variability epoch, while the class $\alpha$ can be fit by IR flickering with frequencies in the range 0.1 to 0.3 Hz, strengthening $\sim$10 s after every X-ray subflare. We interpret these features in the context of the X-ray-emitting accretion disk and IR emission from relativistic jet formation in GRS 1915+105, concluding that the CCF analysis places the origin in a synchrotron-emitting relativistic compact jet at a distance from the compact object of $\sim$0.02AU.
\end{abstract}

\keywords{Microquasars, GRS 1915+105, Jet Formation, QPO, X-ray, RXTE, IR, Palomar}

\section{Introduction}
The Galactic microquasar GRS 1915+105 was discovered as an X-ray transient in 1992 \citep{Castro-Tirado94}, and later revealed itself as the most variable microquasar in the sky. The GRS 1915+105 system contains one of the most massive Galactic stellar mass black holes known to date, initial estimations of 14.0$\pm$4.4 \Msun \citep{Harlaftis04} with more recent values of 10.1$\pm$0.6 \Msun \citep{Steeghs13}. GRS 1915+105 exhibits an extremely high accretion rate, which is likely to account for the high variability observed in the radio, infrared (IR), and X-ray emission of this object. During the 20-year life span of the current outburst state, GRS 1915+105 has exhibited highly-variable X-ray emission that has been divided into 12 different classes depending on count rates and color-color diagram \citep{Belloni00}, three types of different IR flares -- classes A, B, and C -- depending on duration and flux density \citep{Eikenberry01}, and the presence of superluminal radio ejections \citep{Mirabel94}. Simultaneous multi-wavelength observations of GRS 1915+105 have proved the interconnection between those phenomena, showing evidence of a disk-jet coupling \citep{Eikenberry98a,Mirabel98,Fender98,Rothstein05}. 

In particular, \cite{Eikenberry98a} and \cite{Rothstein05} show clear evidence of the connection between the X-ray emitting inner accretion disk and the IR compact jet with the observation of several cycles of simultaneous IR flares and high X-ray variability. \cite{Eikenberry98a} present repeated episodes of simultaneous synchrotron produced class B IR flares ($\sim$100 mJy) and class $\beta$ X-ray variability as the first clear observational evidence of the interaction between the inner disk and the relativistic compact jet in a black hole system. The IR emission appears to be synchrotron in origin \citep{Fender97,Pooley97}, with adiabatic expansion as the dominant cooling mechanism \citep{Mirabel98}, and the observations rule out thermal reprocessing of the X-ray flux on the outer disk and/or the companion star as the origin of the IR flares because of the decoupling of the X-ray and IR excesses at late times in some of the flares. \cite{Eikenberry98a} explain those episodes via a cyclical process where the hot inner accretion disk is emptied by means of a synchrotron-emitting jet ejection event and subsequently replenished by accretion.

\cite{Eikenberry00} explains the IR excess observed after the decoupling of the \cite{Eikenberry98a} IR and X-ray flares and before the return to the quiescent value by associating each X-ray oscillation with a small (5-10 mJy) IR subflare. \cite{Rothstein05} use a similar idea to explain the IR flares observed in repeated episodes of simultaneous class C IR flares ($\sim$40 mJy) and class $\alpha$ X-ray variability. They associate each X-ray oscillation with a simulated 8 mJy Gaussian subflare with variable FWHM, obtaining a simulated overall IR flare (composed of the superposition of these faint subflares) that closely resembles the observed IR light curve.

The study of simultaneous light curves from different energy bands in microquasars is currently the most promising path to extending our knowledge of the various processes happening on microquasars, especially accretion, relativistic ejection, and reprocessing. Techniques such as the cross-correlation function (CCF) provide us with information about the time lags between the different components of the emission (i.e., between the physical processes happening in the objects), thus allowing us to constrain the internal physics of the systems and facilitating the development of more accurate models of the emission of microquasars. Very few studies exist in the literature where the CCF technique has been applied to multi-wavelength observations of microquasars, due both to the difficulty of obtaining simultaneous high time-resolution observations with different instruments, and the difficulty in catching the transient outburst phases of microquasars. Some examples of simultaneous multi-wavelength observations where the CCF technique is applied are the microquasars GX 339-4 \citep{Gandhi08,Gandhi10,Durant11}, SWIFT J1753.5-0127 \citep{Durant08,Durant11}, Sco X-1 \citep{Durant11}, and Cyg X-2 \citep{Durant11}, which were observed in the optical (VLT/ULTRACAM) and X-ray (RXTE/PCA) wavelengths, always during the low/hard state of the microquasars. The CCF technique reveals lags between the optical and the X-ray emission in all those objects -- with the optical leading the X-ray -- this cannot be explained by the reprocessing of the X-ray radiation on the outer disk and/or the surface of the companion star. This implies that microquasar models require a new source of optical emission. However, CCF studies involving IR and X-ray radiation of microquasars were limited until now to the microquasar GX 339-4. \cite{Casella10} present a sub-second time-resolution CCF study of the IR (VLT/ISAAC) and X-ray (RXTE/PCA) emission of GX 339-4 during the highly variable low/hard state. They observe a strong correlation between the IR and the X-ray emission, with the IR lagging the X-ray by 100 ms, thus constraining the Lorentz factor of the jet to $\Gamma$ $>$ 2 and the jet speed to mildly relativistic near its formation region.

In the case of GRS 1915+105, most of the IR emission is blocked by the high extinction toward its location in the Galactic Plane, which reduces its brightness to K $\sim$ 13 mag, and limits the signal-to-noise (S/N) and time resolution of IR observations. In this work, we analyze the behavior of the X-ray to IR CCF during the two sets of observations presented by \cite{Eikenberry98a} and \cite{Rothstein05}. We also use the results of the CCF to compare simulated IR light curves with the observations. Finally, we consider the implications of these results for relativistic jet formation in GRS 1915+105.

\section{Observations and Data Reduction}
We analyze simultaneous X-ray and near-IR observations of GRS 1915+105 on the nights of 1997 August 14-15 and 2002 July 27-28 -- the same data discussed by \cite{Eikenberry98a} and \cite{Rothstein05}, respectively. For brevity's sake, we provide a short summary of the data reduction here and refer the reader to those articles for more details. We obtained IR observations using the Palomar 200-inch telescope and the Cassegrain D-78 near-IR camera in the K (2.2 $\mu$m) band, with X-ray observations obtained using the Proportional Counter Array (PCA) on-board the Rossi X-ray Timing Explorer (RXTE). The 1998 IR data were obtained with 0.1 s time resolution and rebinned down to 1 s in post-processing to improve signal-to-noise, and the 2002 IR data and the X-ray observations were obtained with 1 s time resolution. Absolute timing for the IR observations was provided by a WWVB receiver with $\sim$1 ms accuracy. IR fluxes are calibrated using the nearby Star A (K = 13.3 mag), and dereddened by A$_{K}$ = 3.3 mag to compensate the high Galactic plane absorption \citep{Fender97}.

\begin{figure}
  \centering
    \includegraphics[bb= 75 140 580 610, width=6.5in, scale=1,clip=true]{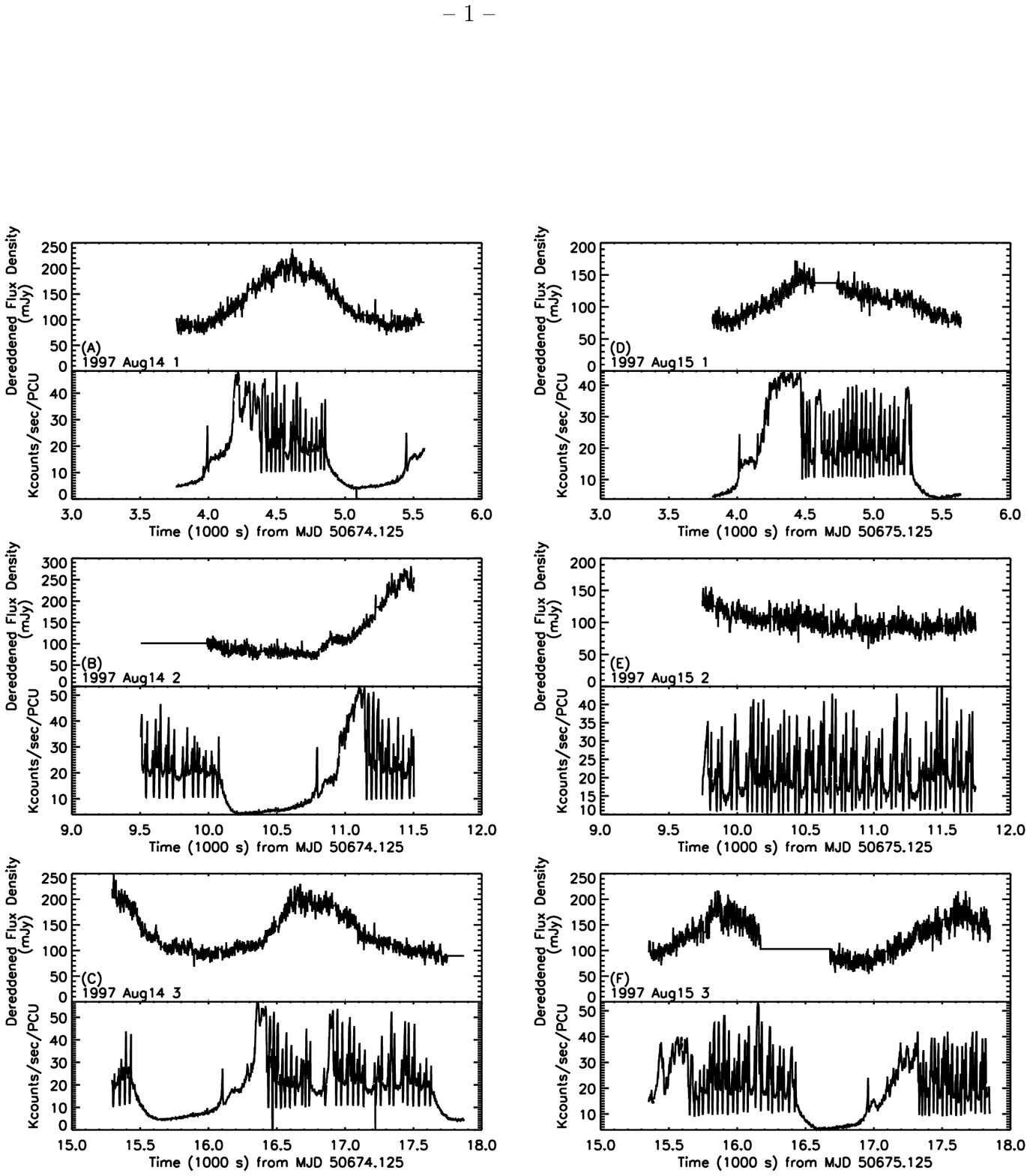}
    \caption{Simultaneous X-ray and IR light curves during the class $\beta$ period (left column: 1997 August 14, right column: 1997 August 15) at 1 s time resolution. Only shown here the portions of the light curves were simultaneous class B IR flares and class $\beta$ high X-ray variability epochs are present.}
    \label{fig:lc_1997}
\end{figure}

\begin{figure}
  \centering
    \includegraphics[width=6.5in, scale=1]{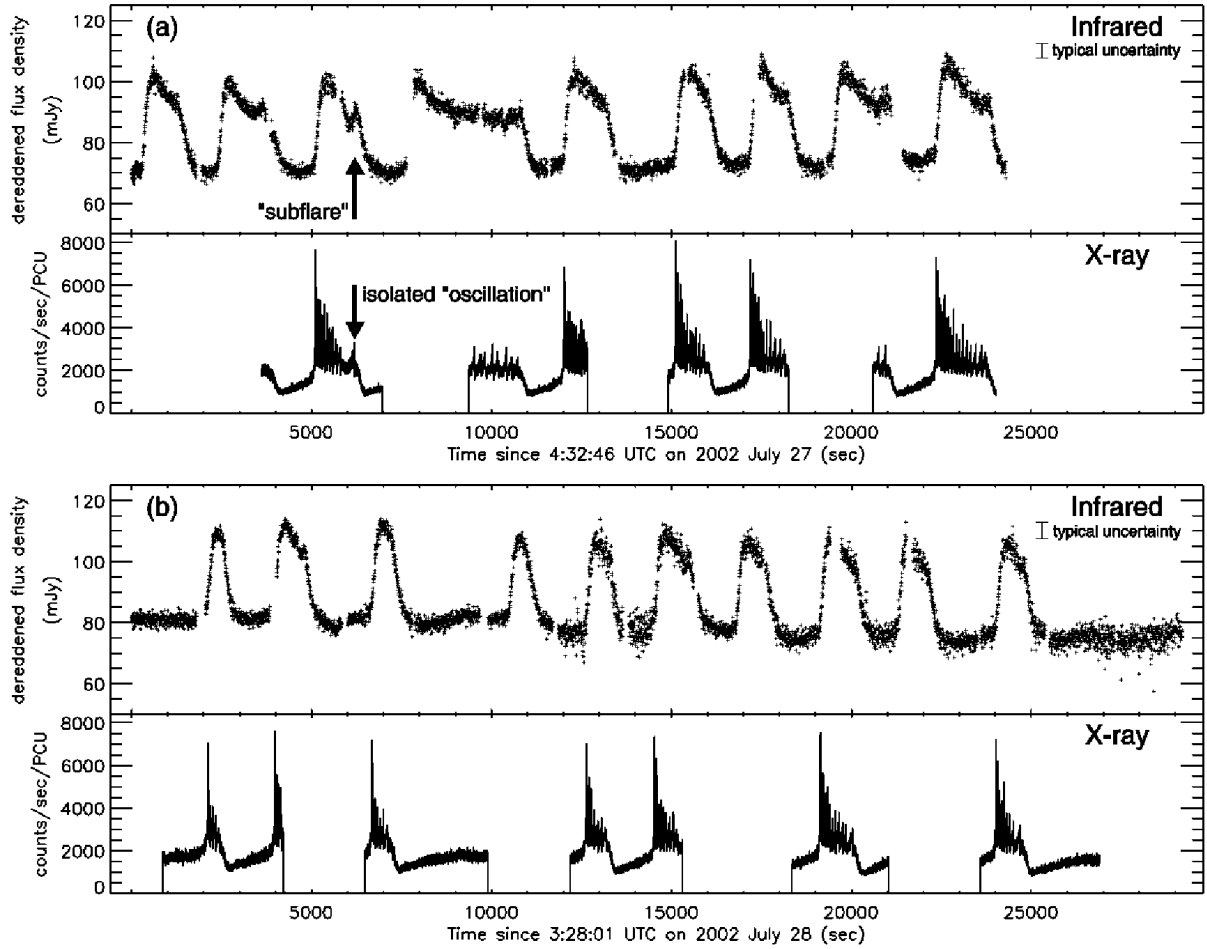}
    \caption{Original simultaneous X-ray and dereddened IR observations during the class $\alpha$ period (upper panel: 2002 July 27, lower panel: 2002 July 28) at 1 s time resolution \cite{Rothstein05}. Several Class C IR flares are simultaneous with class $\alpha$ high X-ray variability epochs.}
    \label{fig:lc_2002}
\end{figure}

Once the \cite{Eikenberry98a} and \cite{Rothstein05} observations have been reduced, we divide them in multiple epochs. Each epoch covers a single full X-ray low/hard state cycle of the light curves -- the X-ray dips -- or a single full high X-ray variability cycle of the light curves -- the X-ray flares. We only analyze those X-ray dips or flares with simultaneous X-ray and IR coverage. Table \ref{tab:time_references} shows the time ranges of each epoch in the references figures (Figs. \ref{fig:lc_1997} \& \ref{fig:lc_2002}).

Also, to confirm the classification of the X-ray observations, we have reduced the original PCA RXTE data and have followed the scheme of \cite{Belloni00} for classifying the X-ray variability. We agree with \cite{Rothstein05} in classifying the observations in 1997 as belonging to class $\beta$, and observations in 2002 as class $\alpha$.

\section{Simultaneous X-ray and IR Light Curves}
\label{simultaneous}
Fig. \ref{fig:lc_1997} shows the portion of the class $\beta$ light curves where simultaneous class B IR flares and high X-ray variability are present, and Fig. \ref{fig:lc_2002} shows the class $\alpha$ X-ray and IR light curves. Both IR fluxes have been dereddened by a factor of 10$^{0.4A_{K}}$, with A$_{K}$ equal to 3.3 mag \citep{Fender97}, to compensate the Galactic plane absorption \citep{Rothstein05}. In each one of the observations, we individually analyze all the portions of the light curves that contain simultaneous coverage of the X-ray low/hard state epochs -- the X-ray dips --, and of the high X-ray variability epochs -- the X-ray flares (Table \ref{tab:time_references}). We divide all the epochs in 60 s bins and calculate the CCF in each bin. We calculate the mean CCF of all consecutive 60 s bins covering each entire long dip and high-variability epoch present in the class $\beta$ and class $\alpha$ periods. 

While this technique shows consistent results for the X-ray dips (which are essentially featureless), it shows inconsistent results for the high-variability epochs because of the presence of the X-ray spikes. The variable periodicity of the X-ray spikes causes the spike to be in a different position in each 60 s time bin.  Therefore, we modified the technique for the high-variability epochs by calculating the mean CCF using 60 s bins centered on each of the X-ray spikes. This gives us a reference for the center of the bins that facilitates the comparison between all bins, although it causes some loss of temporal information when excluding timespans without well-defined spikes. We also tried different bin sizes covering from 20 s up to 120 s obtaining similar results during the low/hard epochs. However, during the high-variability epochs the X-ray spikes are separeted by $\sim$60 s, causing overlap of the CCF of the spikes when choosing bin sizes larger than 60 s, i.e., when more than one spike per bin is present. We also investigate the relative time evolution of the two signals representing the individual CCFs of all 60 s bins versus time for the low/hard state epochs, and versus the spike position -- where spike number 1 is the first well-defined narrow spike -- for the high-variability epochs.

\begin{table*}[tp]%
\scriptsize
\caption{Time reference of all analyzed epochs. We indicate the time range of each epoch in the reference figures as well as the type of variability. The last column indicates the location of the mean CCF plot associated to each epoch.}
\label{tab:time_references}
\begin{center}
{
\begin{tabular}{c c c c c}
\hline
Epoch 		& Variability	& Reference	& Time Range	& Mean CCF	\\
		& Type		& Figure		& (sec)		& Figure		\\
\hline
1997 Aug14 1	& Dip		& Fig.1 A	& 4905-5444	& Fig.3 A	\\
1997 Aug14 2	& Dip		& Fig.1	B	& 10100-10759	& Fig.3 B	\\
1997 Aug14 3	& Dip		& Fig.1	C	& 15475-16074	& Fig.3 C	\\
1997 Aug15 1	& Dip 		& Fig.1 F	& 16430-16969	& -		\\
2002 Jul27 1	& Dip		& Fig.2	A	& 11000-12019	& Fig.3 D	\\
2002 Jul27 2	& Dip		& Fig.2 	A	& 16200-17219	& Fig.3 E	\\
2002 Jul27 3	& Dip		& Fig.2 	A	& 21200-22279	& Fig.3 F	\\
2002 Jul28 1	& Dip		& Fig.2 	B	& 2800-3999	& -		\\
2002 Jul28 2	& Dip		& Fig.2 	B	& 7500-9959	& -		\\
2002 Jul28 3	& Dip		& Fig.2 	B	& 13500-14579	& -		\\
2002 Jul28 4	& Dip		& Fig.2 	B	& 20400-21179	& -		\\
2002 Jul28 5	& Dip		& Fig.2 	B	& 25100-27019	& -		\\
\hline
1997 Aug14 1	& Flare		& Fig.1	A	& 4365-4844	& Fig.4 A	\\
1997 Aug14 2	& Flare		& Fig.1	B	& 11185-11484	& Fig.4 B	\\
1997 Aug14 3	& Flare		& Fig.1	C	& 16435-17634	& Fig.4 C	\\
1997 Aug15 1	& Flare		& Fig.1	D	& 4420-5199	& Fig.4 D	\\
1997 Aug15 2	& Flare		& Fig.1	E	& 9745-11724	& Fig.4 E	\\
1997 Aug15 3	& Flare		& Fig.1	F	& 15650-16429	& Fig.4 F	\\
2002 Jul27 1	& Flare		& Fig.2	A	& 5170-6309	& Fig.5 A	\\
2002 Jul27 2	& Flare		& Fig.2	A	& 12100-12580	& Fig.5 B	\\
2002 Jul27 3	& Flare		& Fig.2	A	& 15150-16109	& Fig.5 C	\\
2002 Jul27 4	& Flare		& Fig.2	A	& 17250-18269	& Fig.5 D	\\
2002 Jul27 5	& Flare		& Fig.2	A	& 22400-23959	& Fig.5 F	\\
2002 Jul28 1	& Flare		& Fig.2	B	& 2270-2749	& -		\\
2002 Jul28 2	& Flare		& Fig.2	B	& 6800-7399	& -		\\
2002 Jul28 3	& Flare		& Fig.2	B	& 12750-13409	& -		\\
2002 Jul28 4	& Flare		& Fig.2	B	& 14650-15429	& -		\\
2002 Jul28 5	& Flare		& Fig.2	B	& 19250-20329	& -		\\
2002 Jul28 6	& Flare		& Fig.2	B	& 24150-24989	& -		\\
\hline 
\end{tabular}
}
\end{center}
\end{table*}

\section{Observed X-ray to IR Cross-correlations}

\subsection{Time-averaged X-ray to IR Cross-correlations}
\label{time-averaged}
\begin{figure}[ht]
  \centering
    \includegraphics[bb= 80 125 555 635, width=6.5in,scale=1.0,clip=true]{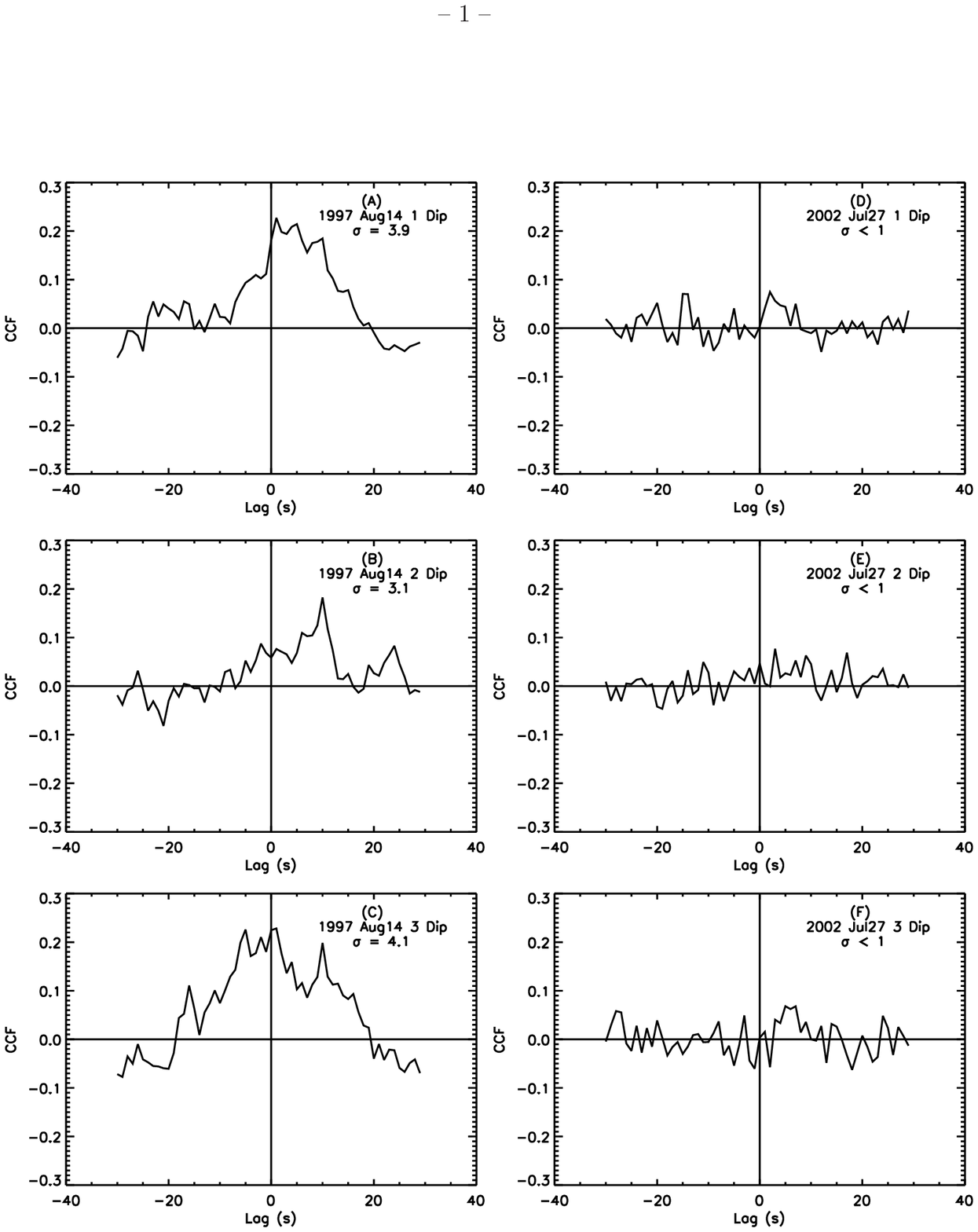}
    \caption{Mean CCF of each of the August 14 class $\beta$ (left column) and July 27 class $\alpha$ (right column) low/hard state epochs shown in Fig. \ref{fig:CCF_dips_3D_14_27}. Class $\alpha$ mean CCFs vary as a white noise with correlations within the $\pm$0.08 range. We use class $\alpha$ mean CCFs as a baseline to evaluate the other CCFs in this paper. Class $\beta$ mean CCFs have correlations $\sim$3 times larger than the baseline.}
    \label{fig:CCF_dips_Aug14_Jul27}
\end{figure}

For a unified comparison of the observations, we calculate the mean of the individual CCFs of all the 60 s bins into which we have divided each epoch (Sec. \ref{simultaneous}). Fig. \ref{fig:CCF_dips_Aug14_Jul27} shows the mean CCF of the class $\beta$ and $\alpha$ low/hard state epochs. The class $\alpha$ mean CCFs resemble white noise with amplitudes in the $\pm$0.08 range, showing no apparent significant correlation. To verify this conclusion, we carried out simple Monte Carlo simulations of the CCFs as follows. For each 60 s bin used in the CCF measurements, we determine the mean and RMS of the IR lightcurve. We then create 10,000 simulated IR light curves with the same mean, and Gaussian noise with the same RMS as the observation. We cross-correlate this simulated light curve with the real X-ray light curve (which has clear, high signal-to-noise variability) to obtain a simulated ``noise'' CCF. We then average these simulated CCFs in the same manner as the real CCFs to obtain 10,000 ``mean noise CCFs''. We then count the number of mean noise CCFs with maximum amplitude greater than the maximum amplitude of the observed CCF for the same epoch, to find the probability that noise could produce a CCF of this amplitude. We find no significant correlations in the $\alpha$ low/hard state observations -- all have confidence level $<$95 $\%$ --, and are generally consistent with pure noise.

The mean CCF plots of the class $\beta$ low/hard state epochs present strong correlations, with a recurring pattern seen in the Fig. \ref{fig:CCF_dips_Aug14_Jul27}. We repeated similar Monte Carlo simulations to those described in the preceding paragraph, and we find that these CCFs have confidence level $>$99.9 $\%$ (equivalent to $>$3 $\sigma$) that they are not due to noise. However, in Sec. \ref{sec:evolution} we show that the strong correlation present in the class $\beta$ low/hard state is present early in each epoch, and is consistent with the overall slow flux decreases from the global pattern shown in Figure 1. Furthermore, if we remove a global linear trend from each 60 s IR lightcurve, the observed CCF structure effectively disappears, and the significance seen in Monte Carlo simulations of the detrended data is $<$2 $\sigma$. Thus, we conclude that there is no significant, relevant correlation seen in the $\beta$ low/hard state (matching the result from the $\alpha$ low/hard state).

\begin{figure}
  \centering
    \includegraphics[bb= 80 125 555 635, width=6.5in,scale=1.0,clip=true]{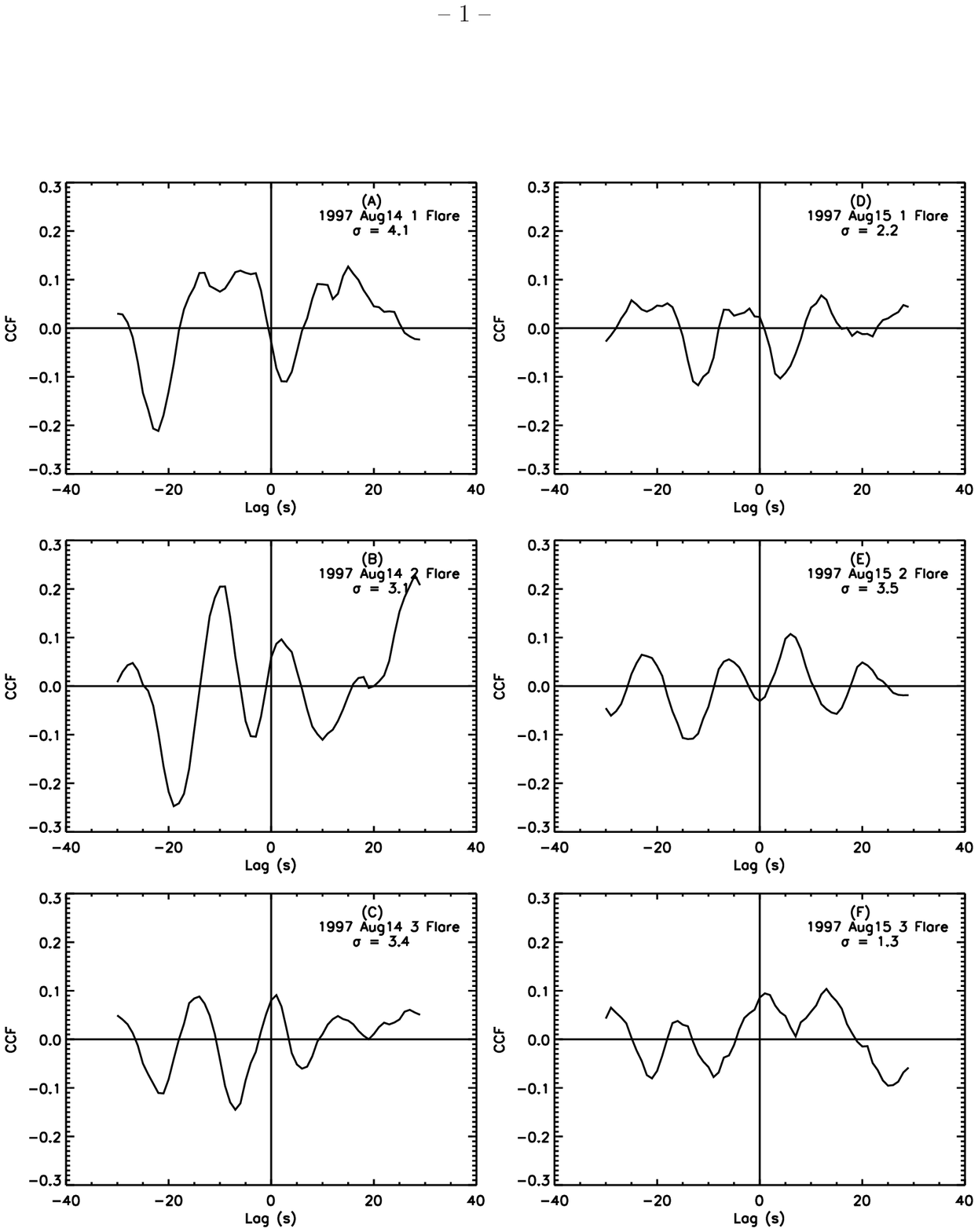}
    \caption{Mean CCF of each of the August 14 (left column) and 15 (right column) class $\beta$ high X-ray variability epochs shown in Fig. \ref{fig:CCF_hearts_3D_14_27}. The mean CCFs present a similar smoothly-varying pattern in all epochs, with maximum correlations between 20 and 120\% larger than the baseline, and with the lag of the larger correlation not constant.}
    \label{fig:CCF_hearts_Aug}
\end{figure}

Figs. \ref{fig:CCF_hearts_Aug} and \ref{fig:CCF_hearts_Jul27} show several of the mean CCFs for the class $\beta$ and $\alpha$ high X-ray variability epochs. The mean CCF plots of the high-variability epochs are similar within each variability type although completely different between them. In other words, the class $\alpha$ high-variability epochs are statistically significant (typical Monte Carlo confidence level $>$ 99.9 $\%$) and all resemble each other, and the class $\beta$ high-variability epochs are also significantly above the noise baseline and all resemble each other, but the class $\alpha$ CCFs are different from the class $\beta$ CCFs. In particular, the $\alpha$ CCFs show a smooth quasi-sinusoidal structure in general, with the strongest average feature being an anticorrelation at typical lags $\sim$ -10 -- -15 s.  The $\beta$ CCFs, on the other hand, show sharper features, both positive and negative, with no apparent consistent lag.

For the $\alpha$ high-variability CCFs, we calculate the measured lag (defined to be the minimum of the anti-correlation) for each of the mean CCFs. We do this by calculating the position-weighted centroid of all CCF values in the lag range from 0 to -25 s. We present the results in Table \ref{tab:centroid}. We find that the lags have a $\pm$1 $\sigma$ range from -7 s to -19 s, with a combined mean/uncertainty of -13$\pm$2 s in the lag. Again, the sense of this lag is such that the IR flare generally lags the X-ray.

We can see from the results of this section that we have succeeded in detecting clear, significant correlations between X-ray and IR features related to jet-producing phases of GRS 1915+105. These are generally significantly fainter in the IR than the previously detected features, and can potentially provide new insight into the jet formation mechanism in this archetypal microquasar. In the following section, we seek to further characterize these new features identified by the CCF technique.

\begin{table*}[tp]%
\scriptsize
\caption{Values of the measured lag obtained calculating the position-weighted centroid of all CCF values in the lag range from 0 to -25 s. The values present a combined mean/uncertainty of -13$\pm$2 s.}
\label{tab:centroid}
\begin{center}
{
\begin{tabular}{c c c c c}
\hline
Epoch 		& Variability	& Mean CCF	& Measured	\\
		& Type		& Figure		& Lag		\\
\hline
2002 Jul27 1	& Flare		& Fig.5 A	& -14.2		\\
2002 Jul27 2	& Flare		& Fig.5 B	& -15.9		\\
2002 Jul27 3	& Flare		& Fig.5 C	& -15.9		\\
2002 Jul27 4	& Flare		& Fig.5 D	& -17.5		\\
2002 Jul27 5	& Flare		& Fig.5 F	& -11.7		\\
2002 Jul28 1	& Flare		& -		& -8.3		\\
2002 Jul28 2	& Flare		& -		& -25.3		\\
2002 Jul28 3	& Flare		& -		& -4.9		\\
2002 Jul28 4	& Flare		& -		& -13.4		\\
2002 Jul28 5	& Flare		& -		& -4.4		\\
2002 Jul28 6	& Flare		& -		& -16.1		\\
\hline 
\end{tabular}
}
\end{center}
\end{table*}

\subsection{CCF Temporal Analysis}
\label{sec:evolution}
The mean CCF method described above has revealed significant general correlations between the X-ray and IR light curves that are worthy of temporal analysis (i.e., to identify possible evolution with time). Thus, we used the individual CCFs of the 60 s bins to study the time variability between the simultaneous X-ray and IR light curves of GRS 1915+105 during the low/hard state and the high X-ray variability epochs. We initially analyzed the evolution of the CCF within the low/hard state epochs present in the class $\beta$ and $\alpha$ periods (Fig. \ref{fig:CCF_dips_3D_14_27}). While the class $\alpha$ low/hard state epochs show no clear evidence of correlations, we observe a small evolution of the CCF during the class $\beta$ low/hard state epochs with a strong correlation in the first 3 bins (180 s) at approximately zero lag that fades with time. Since the correlation is happening when the X-ray and IR fluxes are still globally decreasing, we believe this strong correlation could be contamination due to a remnant of the previous high X-ray variability epochs. This is consistent with ``detrended'' CCFs above, which show that removing long-term linear trends from the IR lightcurve effectively eliminates the statistical significance of the CCFs for the $\beta$ dips. 

Unlike the dips, the high X-ray variability epochs show no CCF resemblance between the class $\beta$ and $\alpha$ epochs. Fig. \ref{fig:CCF_hearts_3D_14_27} shows the evolution of the CCF within the high X-ray variability epochs of both variability types. We are unable to discern any clear temporal evolution in the CCFs in either class. Given that even the mean CCFs typically have statistical significance only at the 3 -- 4 $\sigma$ level, subdividing these detections into multiple (typically $\sim$6 to $\sim$20 individual timebins) would tend to reduce the signal-to-noise below reasonable detection threshholds unless the signal was predominantly concentraed in some subset of the time span.  Thus, we conclude that the observed lack of clear time evolution is consistent with the CCF ``signal'' being more or less evenly distributed with time across each bin include in the mean CCF.

\subsection{The Class $\alpha$ Isolated Subflare}
\label{subsection:isolated}
\begin{figure}
  \centering
    \includegraphics[bb= 80 125 555 635, width=6.5in,scale=1.0,clip=true]{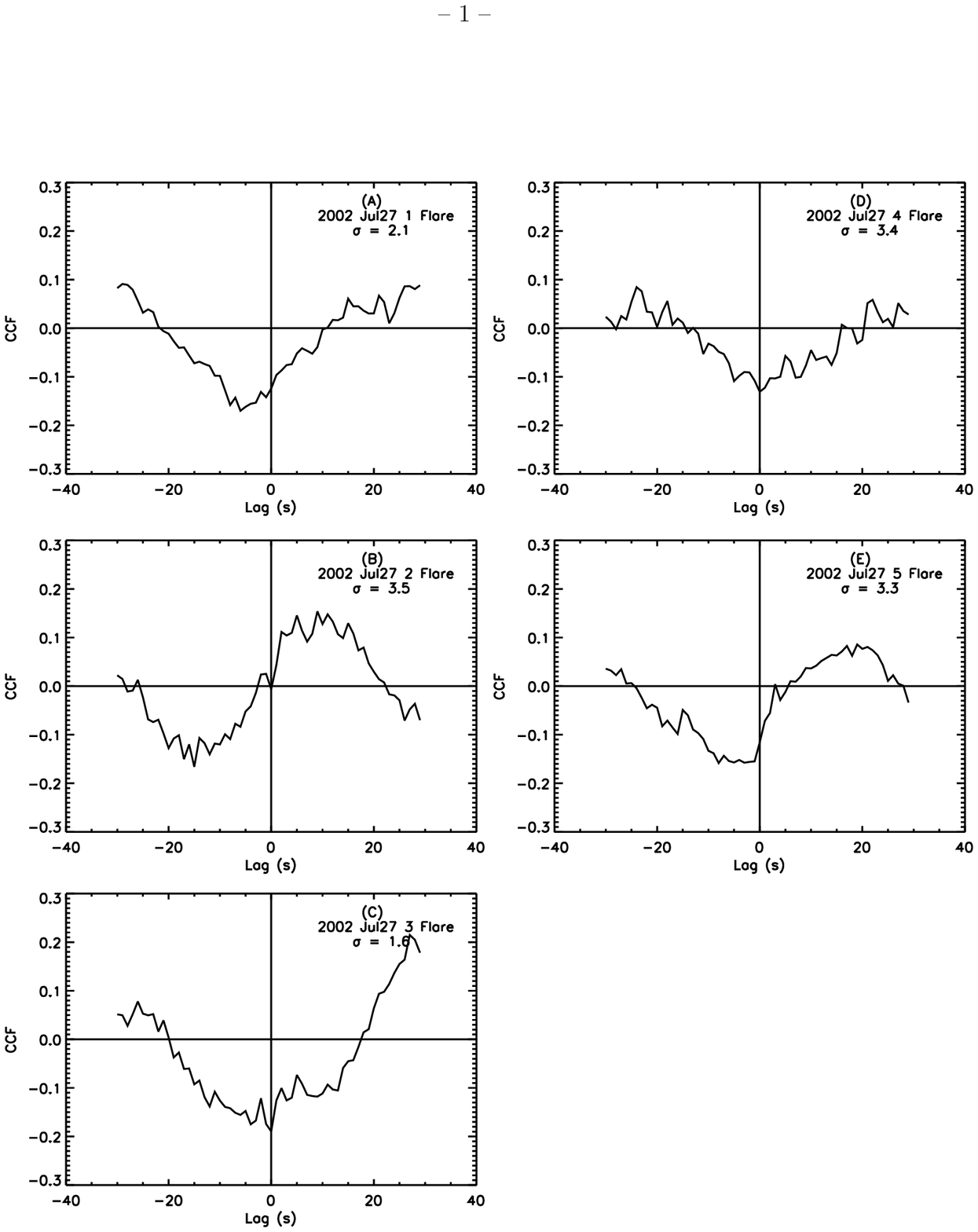}
    \caption{Mean CCF of each of the July 27 class $\alpha$ high X-ray variability epochs shown in Fig. \ref{fig:CCF_hearts_3D_14_27}. The mean CCFs present a similar hardly-varying half-period sinusoidal pattern in all epochs, with maximum correlations between 20 and 150\% larger than the baseline. Maximum correlations are always at positive lags with maximum anticorrelations at negative lags.}
    \label{fig:CCF_hearts_Jul27}
\end{figure}

\begin{figure}
  \centering
    \includegraphics[width=6.5in,scale=1.0]{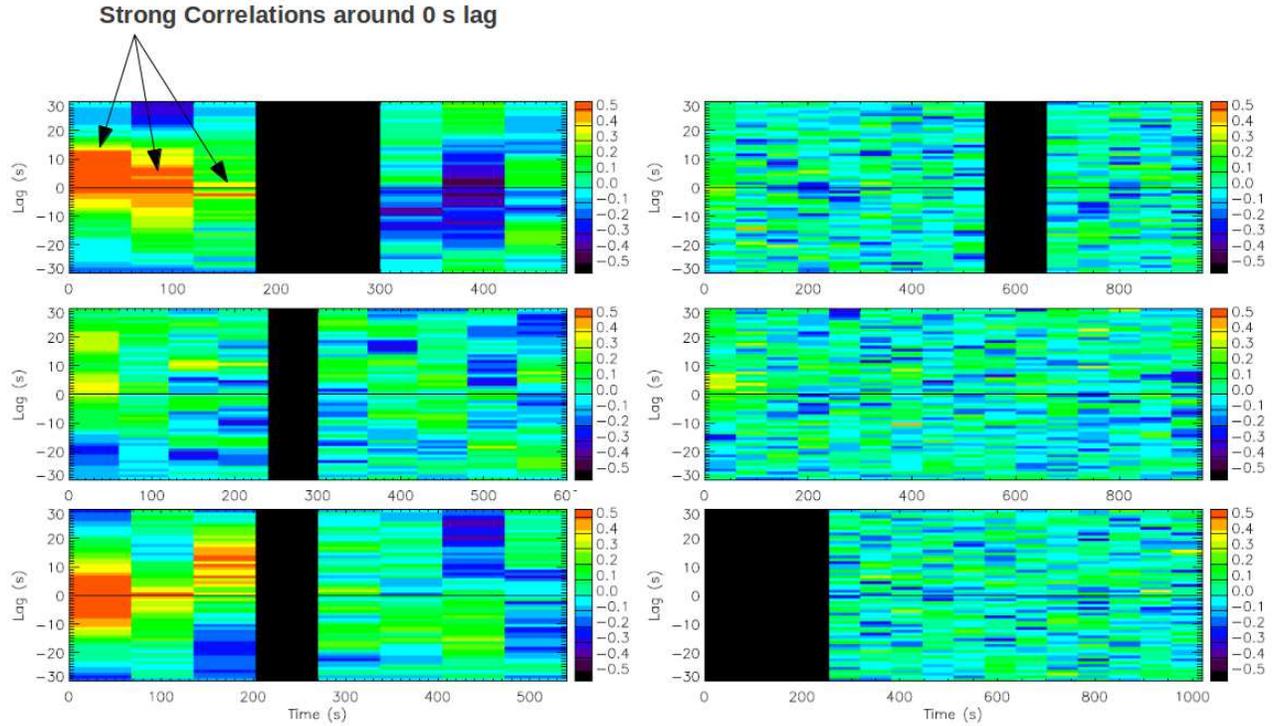}
    \caption{Evolution of the 60 s CCFs during the August 14 class $\beta$ (left column) and July 27 class $\alpha$ (right column) low/hard state epochs. Clear evidence of an evolutionary pattern is seen in the class $\beta$ epochs, with a strong correlation at the beginning of the epochs -- yellow-orange areas below 180 s -- that disappears with time. Despite being in a similar state, the class $\alpha$ epochs do not show such an evolutionary pattern.}
    \label{fig:CCF_dips_3D_14_27}
\end{figure}

\begin{figure}
  \centering
    \includegraphics[bb= 7 105 596 690, width=6.5in,scale=1.0,clip=true]{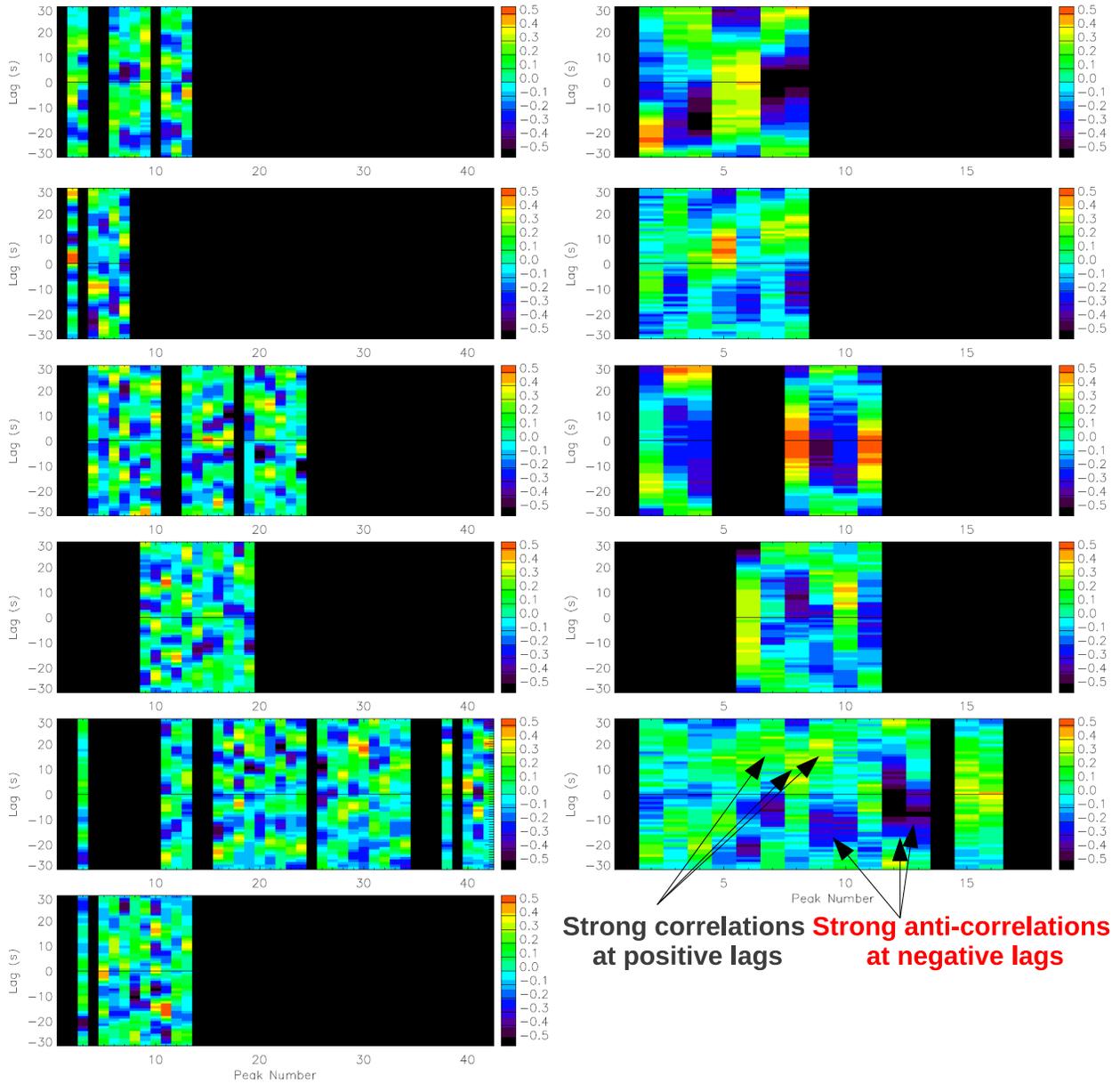}
    \caption{Evolution of the 60 s CCFs centered on the X-ray spikes during the August 14 (three upper panels of left column) and 15 (three lower panels of left column) class $\beta$, and July 27 (right column) class $\alpha$ high X-ray variability epochs. No evolutionary pattern is observed in any of the plots, although the class $\alpha$ CCFs show higher tendency for anti-correlations at negative lags than at positive lags.}
    \label{fig:CCF_hearts_3D_14_27}
\end{figure}

\begin{figure}
  \centering
    \includegraphics[width=6.5in,scale=1.0,clip=true]{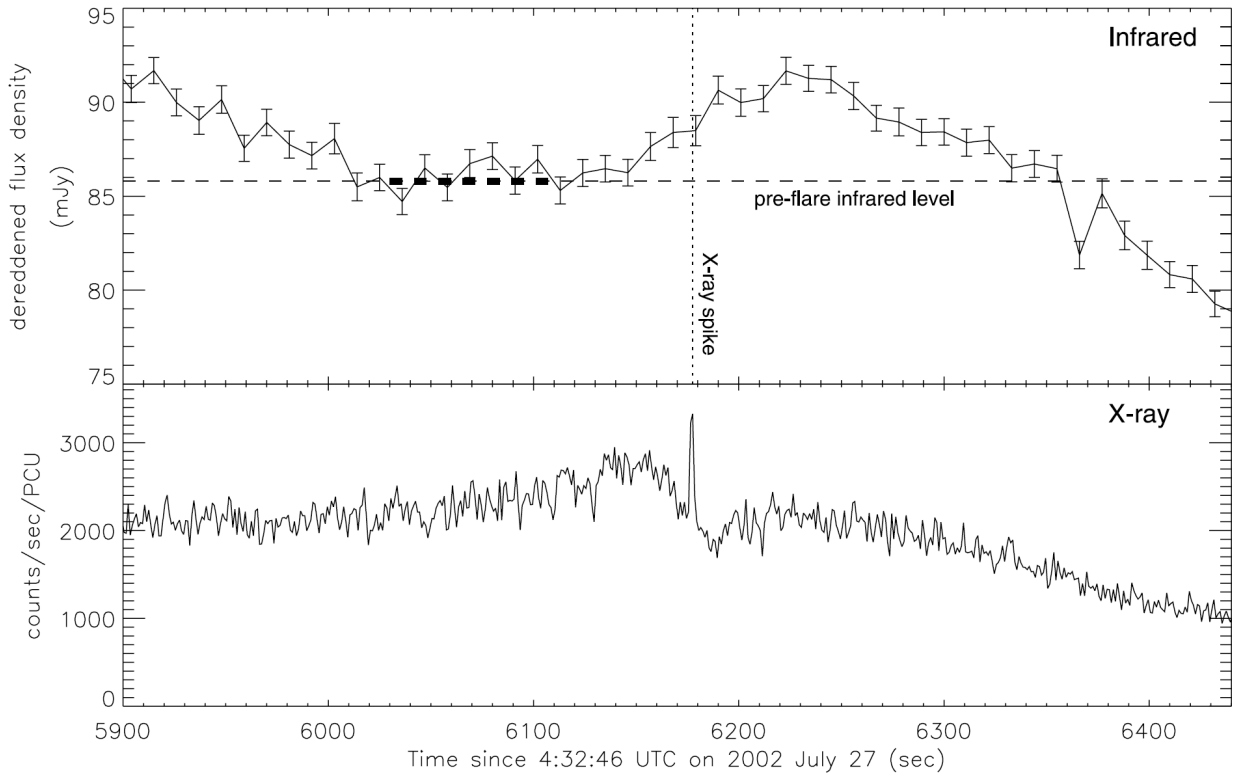}
    \caption{Simultaneous IR and X-ray light curves of the isolated subflare observed during the July 2002 observation \citep{Rothstein05}}
    \label{fig:lc_subflare}
\end{figure}

\begin{figure}
  \centering
    \includegraphics[width=6.5in,scale=1.0,clip=true]{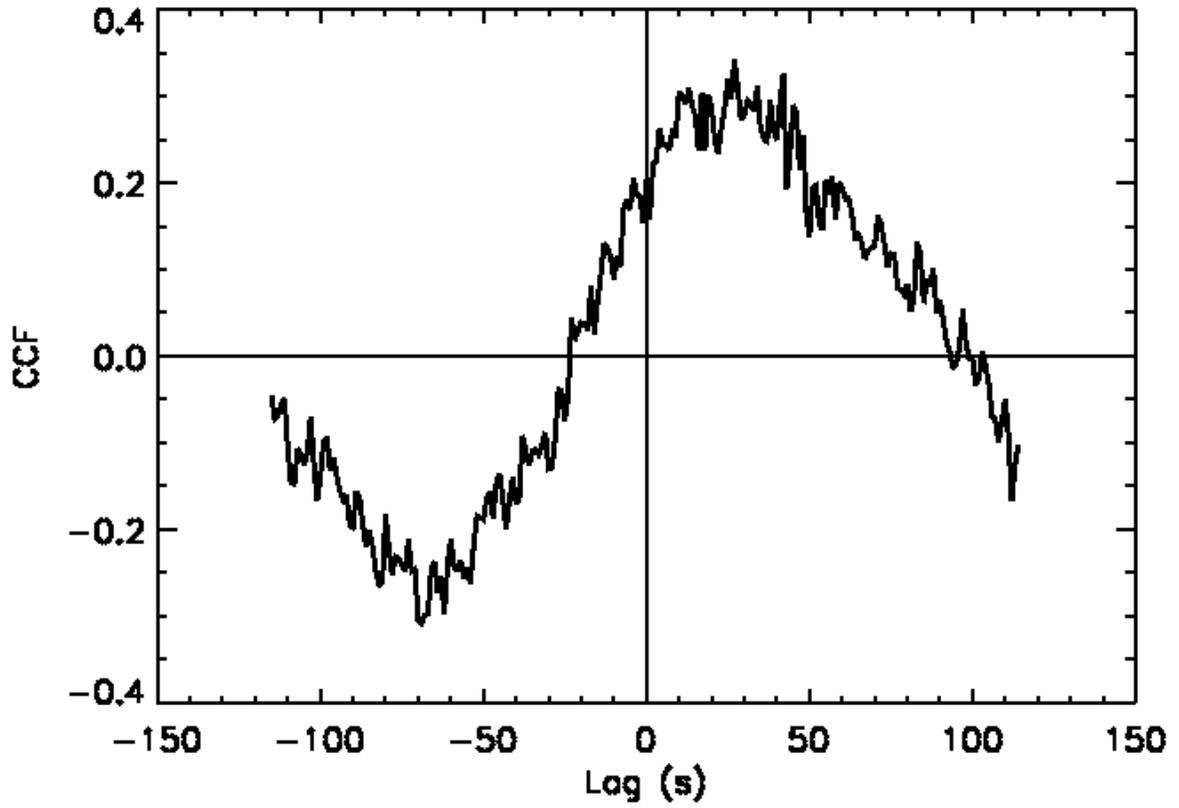}
    \caption{CCF of the isolated subflare observed during the July 2002 observation.}
    \label{fig:ccf_subflare}
\end{figure}

One of the most outstanding features of the class $\alpha$ observations (Fig. \ref{fig:lc_2002}) is the presence of an isolated subflare at the end of a flaring episode. Fig. \ref{fig:lc_subflare} shows the simultaneous X-ray and IR lightcurves for this event \citep{Rothstein05}. The presence of an isolated oscillation was the key factor that led \cite{Rothstein05} to propose the IR flares to be formed as the sum of multiple narrower subflares. For our study, the isolated subflare provides an excellent opportunity for comparison to the other $\alpha$-class analyses presented above. In the Fig. \ref{fig:ccf_subflare} we present the CCF spanning 230 s center on the IR spike. The overall structure of the CCF is remarkably similar to the mean CCFs for the $\alpha$ high-variability epochs shown in Fig. \ref{fig:CCF_hearts_Jul27} above, with a smooth overall quasi-sinusoidal shape and a strong anti-correlation present at negative lags. The primary difference here is that the lag timescale for the isolated flare CCF seems to be $\sim 5$ times longer/slower than we see in the mean CCFs in Fig.\ref{fig:CCF_hearts_Jul27}.

In the isolated subflare case, we can see that the negative lag (indicating that the IR ``lags'' the X-ray emission) comes from the fact that the IR flare -- while it begins {\it before} the primary X-ray spike -- reaches its peak well after the (much faster) X-ray spike.  Thus, the IR lag seems to be more indicative of the emission reaching its peak, rather than which waveband became first.  We therefore assume that this may also be the case for the $\alpha$ high-variability epochs analyzed above.  This distinction, while subtle, may have important ramifications for the origin of this behavior, in that IR-first behaviors would seem to indicate an ``outside-in'' origin for these events (i.e. \cite{Rothstein05}).

\section{Simulated IR Light Curves}
\label{sec:simulation}
\begin{figure}
  \centering
    \includegraphics[bb= 78 118 460 660, width=4.7in,scale=1.0,clip=true]{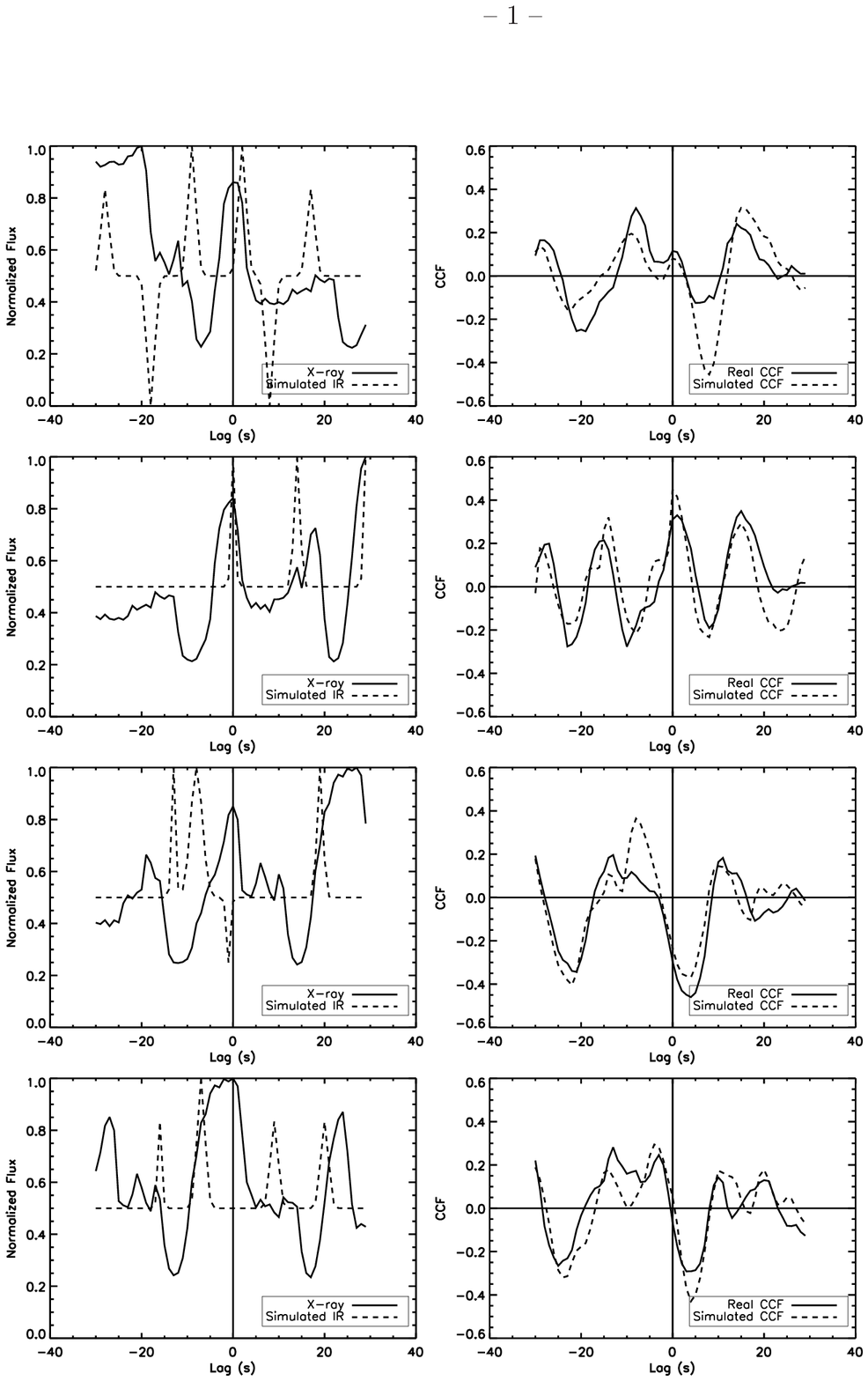}
    \caption{Normalized real X-ray and simulated IR light curves (left column), and real and simulated CCFs (right column) for the spike numbers 2, 3, 6, and 7 of the August 14 class $\beta$ first epoch (top left panel in Fig. \ref{fig:CCF_hearts_3D_14_27}).}
    \label{fig:simulated_Aug14}
\end{figure}

\begin{figure}
  \centering
    \includegraphics[bb= 78 118 460 660, width=4.7in,scale=1.0,clip=true]{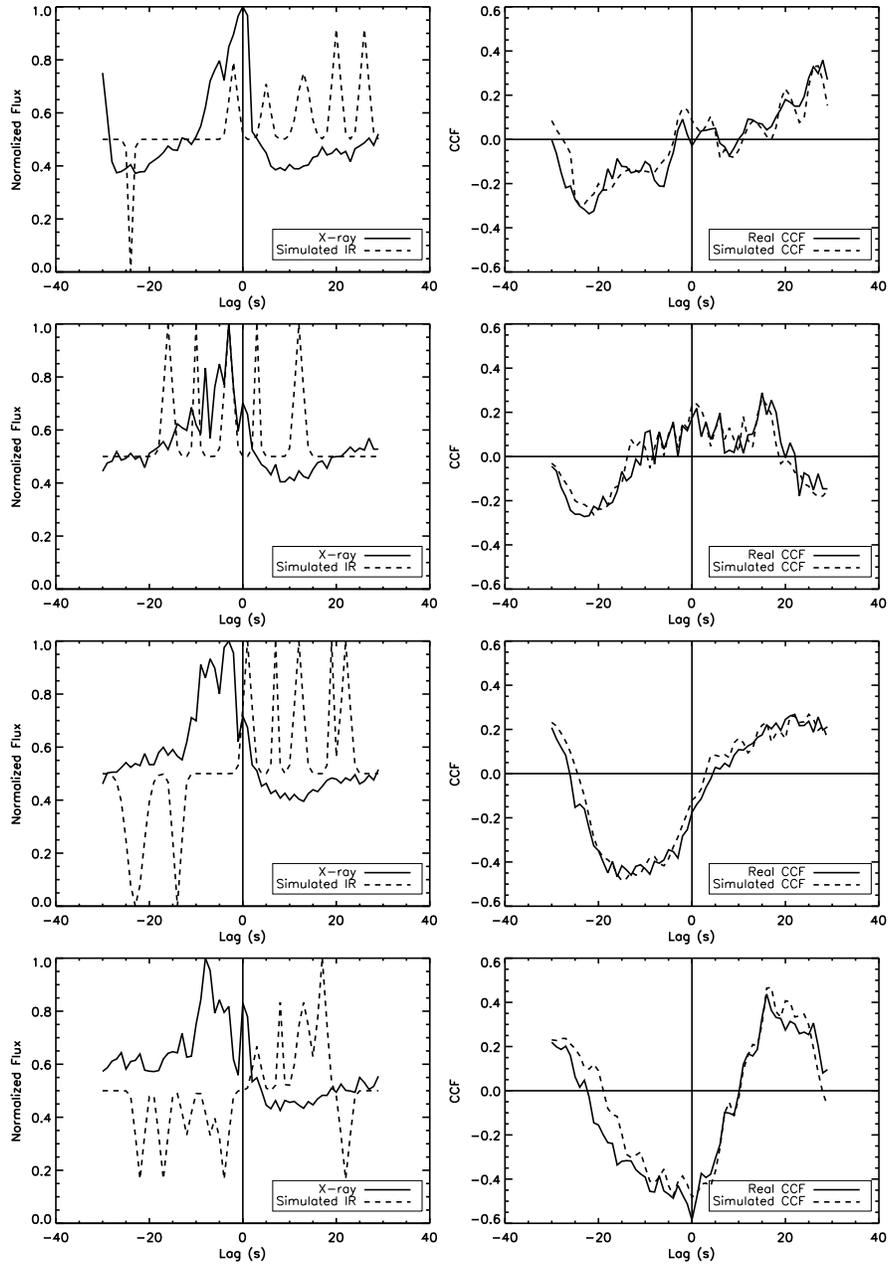}
    \caption{Normalized real X-ray and simulated IR light curves (left column), and real and simulated CCFs (right column) for the spike numbers 2, 3, 4, and 7 of the July 27 class $\alpha$ first epoch (top right panel in Fig. \ref{fig:CCF_hearts_3D_14_27}).}
    \label{fig:simulated_Jul27}
\end{figure}

\begin{figure}
  \centering
    \includegraphics[width=6.5in,scale=1.0]{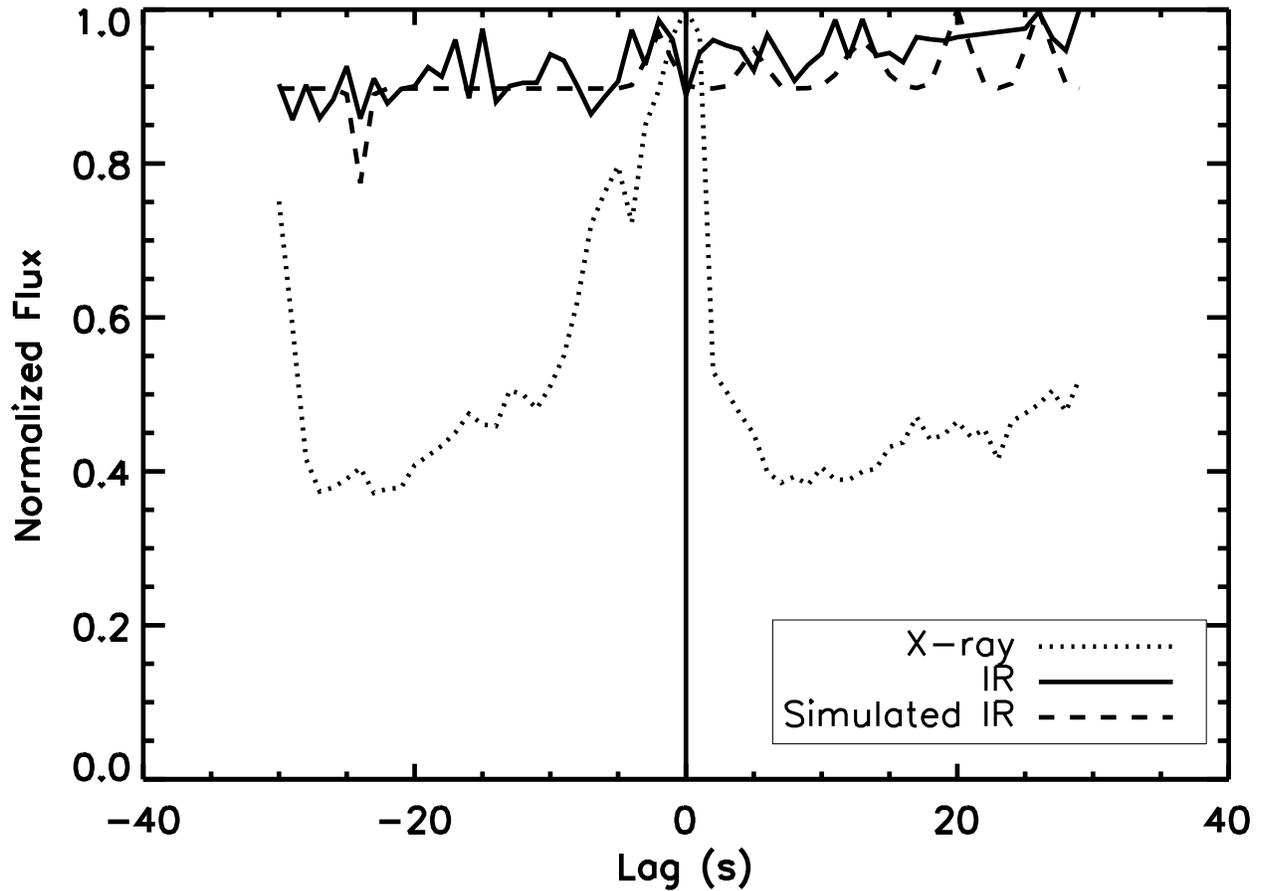}
    \caption{Normalized real X-ray, real IR, and simulated IR light curves for the spike number 2 of the July 27 class $\alpha$ first epoch (top right panel in Fig. \ref{fig:simulated_Jul27}). For a better visualization of the simulated IR light curves, we have added to the group of Gaussians the mean value of the real IR light curve (the flare baseline), and also matched the rms variability of the group of Gaussian with the rms variability of the real IR without varying the shape of the simulated IR. The real and simulated IR light curves present a similar shape.}
    \label{fig:simulated_baseline_Jul27}
\end{figure}

The CCF study indicates that a strong correlation exists between the X-ray and IR fluxes during the class $\alpha$ high-variability epochs. Although, these results are very interesting for constraining the origin of the IR emission, their mere presence does not give full information on the processes happening in the system. In order to improve our understanding of these processes, we use the results of the X-ray to IR CCF to compare the real 60 s IR light curves centered on the X-ray spikes (Sec. \ref{simultaneous}) with simulated 60 s IR light curves. We focus on IR simulations due to the lower S/N here, implying that the CCFs may be picking up behavior not obvious in the IR lightcurves. We created a simulated 60 s IR light curve for each of the high X-ray variability spikes of both class $\beta$ and $\alpha$ epochs, and use the Kolmogorov-Smirnov (KS) test to compare the results of the X-ray to simulated IR CCF with the results of the X-ray to real IR CCF. This allows us to test the hypothesis that the simulated CCF has both the right amplitude and the correct shape as well. The simulated 60 s IR light curves are composed of a variable number of Gaussian subflares with relative amplitudes ranging between 0 and 1 and with FWHM from 0.5 to 3 s. We manually modify the parameters of the Gaussians -- center, amplitude, and FWHM -- based on the results of the KS test. Because we are comparing the relative shape of the light curves, the amplitude of the simulated signals is baseline-subtracted, leaving only the relative flux variation between the peaks. Figs. \ref{fig:simulated_Aug14} and \ref{fig:simulated_Jul27} show the normalized 60 s X-ray and simulated IR light curves (left columns) and the X-ray to real and simulated CCF (right columns) for some of the class $\beta$ and $\alpha$ high X-ray variability spikes, respectively. The KS test indicates that the probability of the two CCF being similar for each individual spike is always higher than 78 \% and in some cases -- especially in the class $\alpha$ spikes -- beyond 98 \% (Table \ref{tab:probability}). Despite having high probabilities, these results are not statistically conclusive by themselves to confirm the simulated light curves as real and unique. To better visualize the simulations, in Fig. \ref{fig:simulated_baseline_Jul27} we show one of the simulated IR light curves from Fig. \ref{fig:simulated_Jul27}. We have re-added the flare baseline of the real IR light curve to the group of Gaussians, which we have scaled to match the rms variability of the real IR light curve. 

The analyses of the isolated subflare (Subsec. \ref{subsection:isolated}) show the presence of broader Gaussians in the IR emission than the ones used for our simulations. Based on that we also attempted to match the observed CCFs of both $\beta$ and $\alpha$ spikes with single Gaussian IR flares. For the $\beta$ spikes we find no fittable model, while for the $\alpha$ spikes we find consistent results with Gaussians of $\sim 25s$ FWHM in the IR. However, the CCFs do not match quite as closely as the narrow Gaussian models (KS test results of $\sim$20 \%), but this model uses fewer degrees of freedom. In short, while the $\alpha$ class CCFs are consistent with the faster multiple IR flares, with the current data we cannot conclusively distinguish from a single slower flare either. However, the lack of a suitable model of a single broad Gaussian for the $\beta$ spikes supports the idea of a multiple narrow Gaussian simulation for the $\alpha$ spikes, reinforcing thus the results showed by the KS test. Based on these evidences, we focus our analyses on the multiple narrow Gaussians simulations of both classes.

To determine the statistical significance of the simulated IR light curves we perform a Monte Carlo simulation of the simulated IR light curves using the same group of narrow Gaussians noted above in each case but allowing random values of the center of the Gaussians. We perform 1000 realizations for each 60 s bin and found that for both classes spikes $<$1 \% exceed the manual CCF KS values. Table \ref{tab:probability} shows the results of the Monte Carlo simulations. The Monte Carlo simulation indicates that the probability of having KS test values equal or higher than the one obtained with our simulated light curves when the individual Gaussians are randomly distributed is around 3 $\sigma$ level of statistical significance for both classes. Therefore, assuming all Gaussian subflares within each epoch originate from the same phenomenon, the overall significances of the simulated light curves during the class $\beta$ and $\alpha$, considering only the four example for each class presented here, become $\sim$6 $\sigma$. From these results, we conclude that the simulated IR light curves that we have created to reproduce the IR emission of GRS 1915+105 during the X-ray class $\beta$ and the X-ray class $\alpha$ epochs are statistically significant, indicating the presence of a flickering IR emission in GRS 1915+105 during both high X-ray variability epochs.

\begin{table*}[tp]%
\caption[IR simulated light curves: KS test and Monte Carlo simulation results]{KS test and Monte Carlo simulation results for the CCFs of the class $\beta$ and $\alpha$ high X-ray variability spikes shown in Figs. \ref{fig:simulated_Aug14} \& \ref{fig:simulated_Jul27}.}
\label{tab:probability}
\begin{center}
{
\begin{tabular}{l l l}
\hline
Spike 	& KS test	& Monte Carlo	\\
	& (\%)	& (\%)  		\\
\hline
Aug 14 - 2  & 91.05 & 1.1 \\
Aug 14 - 3  & 78.37 & 0.8 \\
Aug 14 - 6  & 91.05 & 0.1 \\
Aug 14 - 7  & 91.05 & 0.1  \\
Jul 27  - 2  & 98.08 & 0.6 \\
Jul 27  - 3  & 98.08 & 1.4 \\
Jul 27  - 4  & 91.05 & 0.1 \\
Jul 27  - 7  & 78.37 & 0.1 \\
\hline 
\end{tabular}
}
\end{center}
\end{table*}

\begin{table*}[tp]%
\caption[IR simulated light curves parameters]{IR simulated light curves parameters: mean frequency of the IR flickering; ratio of the overall flux at positive lags respect the overall flux at negative lags for the class $\beta$ high X-ray variability spikes shown in \ref{fig:simulated_Jul27}; lag between the X-ray spike and the centroid of the IR excess; and lag between the beginning of the X-ray rise and the first IR subflare of the excess.}
\label{tab:simulated}
\begin{center}
{
\begin{tabular}{l l l l l}
\hline
Spike 	& Frequency	& Post/Neg Lag Flux	& IR Peak Lag 	& IR Rise Lag	\\
	& (Hz)		& (\%)  			& (s)			& (s)	\\
\hline
Jul 27  - 2  & 7$\pm$1 & +24	& 11	& 7 \\

Jul 27  - 3  & 6$\pm$1 & -4 	& 1	& 3 \\
Jul 27  - 4  & 6$\pm$2 & +67 	& 15	& 12 \\
Jul 27  - 7  & 5$\pm$2 & +45 	& 19	& 13 \\
\hline 
\end{tabular}
}
\end{center}
\end{table*}

Thus our analysis of the simulated IR light curves show that the class $\beta$ IR light curves are composed of a group of independent subflares vary around the X-ray emission, producing variable lag structure. Meanwhile the class $\alpha$ simulated IR light curves are consistent in CCF and are composed of quasi-periodic Gaussian subflares with steady 0.1-0.3 Hz frequencies within each 60 s bin (Table \ref{tab:simulated}) (or less likely of a single Gaussian with $\sim$ 25s FWHM). These class $\alpha$ narrow Gaussian simulated IR light curves show, in most cases, increased IR flux after the X-ray spikes, indicating an interconnection between the physical processes happening in the hot inner accretion disk and the IR emission. These results confirm the existence of quasi-periodic oscillations in the IR emission of GRS 1915+105 during the class C IR flares simultaneous with X-ray class $\alpha$ variability \citep{Rothstein05}.

The CCF study establishes a strong relation between the class $\alpha$ simulated IR light curves and the individual X-rays spikes associated with the accretion disk. Assuming the origin of the IR flickering in a relativistic IR jet and taking into consideration that the overall IR flare can be reproduced as a superimposition of small subflares \citep{Rothstein05}, it is straightforward to conclude that the overall class C IR flares simultaneous with X-ray class $\alpha$ variability also have an origin in the outflow events. \cite{Eikenberry98a} previously demonstrated similar linkage to jet production for the class $\beta$ IR flares. However, we also see that the $\alpha$ subflares seem to be far more consistent in the behavior and correlations than the $\beta$ subflares. The class $\alpha$ simulated signals indicate a well-organized jet where the flickering shows a quasi-steady repetition closely linked to the X-ray emission, typical of a jet strongly correlated with the accretion processes. Meanwhile, the class $\beta$ simulated signals show an unorganized jet formed by stochastic processes with loosely linked with the X-ray emission, more typical of a ``noisy" formation process. 

Radio observations of GRS 1915+105 during a low/hard state place the origin of the radio synchrotron emission at a distance of $\sim$50 AU from the accretion disk \citep{Dhawan00,Klein-Wolt02}. \cite{Eikenberry08} place the origin of a $\sim$25 s timescale IR variability at a distance $\leq$2.5 AU, with a jet diameter of $\sim$0.05 AU. Using the same approximation and considering that the IR flickering observed in the simulated $\alpha$ class high-variability IR light curves is between 3 and 8 times faster than any previously observed IR variability in GRS 1915+105 \citep{Eikenberry08}, we can place the source of production of the IR flickering located between 0.3 and 0.8 AU from the compact object, and with a maximum diameter of the plasmoid blobs between 0.006 and 0.016 AU. However, the simulated IR light curves also show a mean time lag between the X-ray spike and the peak of the IR excess of $\sim$13 s -- similar to the mean lag calculate on the mean CCF (Subsec. \ref{time-averaged}) -- and a mean time lags between the beginning of the X-ray rise and the beginning of the first subflare of the excess of $\sim$10 s (Table \ref{tab:simulated}). A delay of 10 s between the X-ray and IR emissions indicates a maximum distance between the accretion disk and the IR launch site of $\sim$0.02 AU at a jet speed near $c$, i.e., $\sim$50 times shorter than expected from the IR flickering. The short distance of the launch site and the lower limit of 0.006 AU for the diameter of the plasmoid blobs obtained from the 3 s flickering indicate that we are observing physical phenomena produced very close to the origin of the IR plasmoid ejection.

We have demonstrated that both measurements of the time lags between the X-ray and IR emission indicate an unambiguous time delay between the X-ray emission and the plasmoid ejections in contradiction with the hypothesis of an ``outside-in" process presented by \cite{Eikenberry00}. Also, the distance obtained for the origin of the IR emission is $\sim$15 times shorter than the accretion disk outer radius of GRS 1915+105 -- $\sim$0.3 AU -- \citep{Rau03}, definitely confirming the inconsistency of our results with the reprocessing in the outer disk and/or companion star as origin of the IR emission. Reprocessing could still be possible in a ``lump" in the inner disk, although the presence of the narrow IR pulses contradicts that explanation.

\section{Summary}
We perform an analysis of the accretion and ejection processes of GRS 1915+105 using simultaneous X-ray and IR light curves during the low/hard state and the high-variability epochs of two X-ray class $\alpha$ and $\beta$ periods. We show that each state produces a particular mean CCF plot that could be used for classification of the states. The CCF study also shows a weak or null interaction between the X-ray and IR fluxes during the low/hard state epochs and the X-ray class $\beta$ epochs, and a strong correlation between them during the X-ray class $\alpha$ epochs. We study the evolution of the CCF during all the epochs and find some resemblance between the low/hard state epochs, accentuating the similarity between the two states despite been in different X-ray classes, and none during the high X-ray variability epochs. Although these results show that the IR emission during the X-ray class $\beta$ epoch is independent from the X-ray emission, they also show some degree of repeatability in the IR emission that could be more easily associated with jet emission than with reprocessing. During the last state analyzed, the class $\alpha$ high X-ray variability epochs, the CCF study shows a strong interconnection between the X-ray and IR light curves, with the X-ray leading the IR emission. We use simulated IR light curves to present evidence of flickering IR ejection events, with frequencies in the range 0.1 to 0.3 Hz, strongly correlated with the X-ray class $\alpha$ variability. The presence of a modulation in the range 3 to 10 s in the IR emission of GRS 1915+105 is in agreement with the $\sim$8 s timescale flux variations found by \cite{Eikenberry08} and are difficult to explain as reprocessing of the X-ray emission in the outer part of the accretion disk and/or companion star. We propose instead that the origin of the IR emission is a synchrotron-emitting compact jet, consistent with several previous authors. The delay between the X-ray and IR light curves indicates a location of the IR launch site very close to the compact object at $<$ 0.02 AU, and with an upper limit of the diameter of the plasmoid blobs between 0.006 and 0.016 AU.

\acknowledgments
The authors wish to thank D. Rothstein for sharing his observational data with us, and M. Durant, and R. Bandyopadhyay for helpful discussions of these results. The authors also thank the anonymous referee for helpful suggestions. The authors acknowledge the support of the University of Florida Alumni Fellowship and the NSF grant AST-0807687. 

\end{document}